\def\mass#1{${\mathrm{#1\,M}_\odot}$}
\def\mmass#1{{\mathrm{#1\,M}_\odot}}
\def\lsun#1{${\mathrm{#1\,L}_\odot}$}
\def\mlsun#1{{\mathrm{#1\,L}_\odot}}
\def\rsun#1{${\mathrm{#1\,R}_\odot}$}
\def\mrsun#1{{\mathrm{#1\,R}_\odot}}
\def\chem#1#2{$\mathrm{^{#2}\kern-0.8pt#1}$}
\def\mchem#1#2{\mathrm{^{#2}\kern-0.8pt#1}}
\def\reac#1#2#3#4#5#6{$\mathrm{\,^{#2}\kern-0.8pt{#1}\,({#3}\,,{#4})\,{}^{#6}\kern-0.8pt{#5}\,}$}
\def\betap#1#2#3#4{$\mathrm{\,^{#2}\kern-0.8pt{#1}\,(\beta^+)\,{}^{#4}\kern-0.8pt{#3}\,}$}
\def\betam#1#2#3#4{$\mathrm{\,^{#2}\kern-0.8pt{#1}\,(\beta^-)\,{}^{#4}\kern-0.8pt{#3}\,}$}
\def\reacbp#1#2#3#4#5#6#7#8{$\mathrm{\,^{#2}\kern-0.8pt{#1}\,({#3}\,,{#4})\,{}^{#6}\kern-0.8pt{#5}\,(\beta^+)\,{}^{#8}\kern-0.8pt{#7}\,}$}
\def\reacbm#1#2#3#4#5#6#7#8{$\mathrm{\,^{#2}\kern-0.8pt{#1}\,({#3}\,,{#4})\,{}^{#6}\kern-0.8pt{#5}\,(\beta^-)\,{}^{#8}\kern-0.8pt{#7}\,}$}
\def\simgr{\mathbin{\;\raise1pt\hbox{$>$}\kern-8pt\lower3pt\hbox{$\sim$}\;}}
\def\simlr{\mathbin{\;\raise1pt\hbox{$<$}\kern-8pt\lower3pt\hbox{$\sim$}\;}}

\documentclass{aa}
\usepackage{graphics}
\begin{document}

\thesaurus{... }

\title{On the third dredge-up phenomenon in asymptotic giant branch stars}
 
\author{N. Mowlavi}
                                

\institute{Geneva Observatory, CH-1290 Sauverny, Switzerland}

\date{Received date; accepted date }

\maketitle

\begin {abstract}
The third dredge-up phenomenon in asymptotic giant branch (AGB) stars is
analyzed through evolutionary model calculations of a \mass{3}, solar
metallicity star. The Schwarzschild criterion is used to test the stability of a
given layer against convection, and the calculations are performed either
with or without extra-mixing below the convective envelope. Based on these
calculations, several questions are addressed regarding the occurrence of the third
dredge-up in AGB star models, the laws governing that phenomenon, and some of its
implications on the structural and chemical evolution of those stars.

The use of the Schwarzschild criterion {\it without} extra-mixing of any sort is
shown to lead to unphysical afterpulse models which prevent the occurrence of third
dredge-up. Model calculations of a \mass{3} star using no extra-mixing confirm the
failure to obtain dredge-up in those conditions. That conclusion is found to be
independent of the mixing length parameter, stellar mass, or numerical accuracy
of the models.

  Model calculations performed on selected afterpulses of the \mass{3} star, but
with extra-mixing
(using a decreasing bubble velocity field in the radiative layers and a diffusion
algorithm for the mixing of the chemical elements),
lead to efficient dredge-ups at a rate of
\mass{10^{-5}-10^{-4}}/yr. Test calculations using different extra-mixing extents
and efficiencies reveal that the dredge-up predictions
are rather insensitive to those extra-mixing parameters. This important
conclusion is understood by analyzing the physics involved in the dredge-up
process. It is shown that the dredge-up rate is determined by the thermal
relaxation time-scale of the envelope as C-rich matter is added from the core
into the envelope. The dredge-up predictions are, however, expected to depend on the
convection prescription {\it in} the envelope.

  Linear relations both between the dredge-up rate and the core mass $M_c$
and between the dredge-up efficiency $\lambda$ and $M_c$ are predicted by
the model calculations. Those linear relations are expected to still
hold when the feedback of the dredge-ups on the AGB evolution
is taken into account. They predict the dredge-up efficiency to level off at
unity during the AGB evolution, at which point the core mass remains constant from one pulse
to the next. The core mass is concomitantly predicted to evolve towards an asymptotic value.
The existence of such an asymptotic core mass naturally provides an upper limit to the mass of
the white dwarf remnant, and helps to constrain the initial-final mass of white dwarfs.

  Synthetic calculations taking into account the dredge-up laws obtained from the
full AGB model calculations  predict a continuous increase of the stellar luminosity
$L$ with time, contrary to the predicted behavior of $M_c$ and $\lambda$. This results from
an adopted dependence of $L$ on both $M_c$ and the radius $R_c$ of the H-depleted core
of the form $L\propto M_c^2/R_c$. As a result of this increase of $L$ with time,
the initial-final mass relation can further be
constrained if mass loss is taken into account.  If, for example, a
superwind is assumed to eject all the remaining envelope of the \mass{3} star
at \lsun{L=15000}, then
the mass of the white dwarf remnant is predicted to be \mass{0.66},
instead of \mass{0.73} predicted by models without dredge-up.

  Finally, the synthetic calculations predict the formation of a \mass{3} carbon star
after about 20 pulses experiencing dredge-up. Taking into account the fact that
the luminosity decreases by a factor of two during about 20\% of the interpulse
phases, such a \mass{3} carbon star could be observed at luminosities as low as \lsun{7500}.

\end{abstract}

\section{Introduction}
\label{Sect:introduction}

  The third dredge-up (3DUP) phenomenon is the process by which ashes from {\it
helium} burning nucleosynthesis are transported from the interior to the surface
of low- and intermediate-mass stars ($M\simlr \mmass{7}$)
during the asymptotic giant branch (AGB) phase (see, e.g., Mowlavi 1998a for a
general introduction to the structural and chemical evolution of AGB stars).
It contrasts with the first and second dredge-ups which
bring to the surface only
ashes from hydrogen burning. The signature of those 3DUP events is
observable in the specific chemical composition characterizing many AGB
stars as compared to those of red giant stars (see, for example, Mowlavi 1998b).
In particular, the high C/O ratios
observed in MS, S, SC (C/O$\simlr$1 in number) and C (C/O$>$1) stars are
explained as resulting from an increase in the carbon abundance due to the 3DUP
(solar C/O$\simeq$0.42).
Similarly, the overabundances (as compared to those in red giants) of fluorine and
s-process elements (Jorissen et al. 1992) attest to the operation of dredge-up in AGB stars.

  Stellar models show that thermal instabilities (called pulses) develop
periodically in the He-burning shell of AGB stars (Schwarzschild and H\"arm 1965), leading to
structural readjustments which provide an adequate stage for the operation of the
3DUP (Iben 1975, Sugimoto and Nomoto 1975). Unfortunately, model predictions
available in the literature do not agree on the characteristics of the 3DUP, such
as the pulse number at which it begins to operate along the AGB or the extent of envelope penetration
in the C-rich layers.
The problem is of particular relevance when confronting chemical abundances
observed at the surface of AGB stars to model predictions.
For example, the famous `carbon star mystery' put forward by Iben (1981) about
the luminosity function of C stars in the Magellanic Clouds revealed, among other
things, the difficulty
of models to explain the low luminosity C stars (see Mowlavi 1998c for
the present state on that question). The few -- but isolated -- successes of AGB
models in obtaining low-luminosity carbon star models in the late 80's suggested
that the solution to the carbon star mystery resides in the use of new radiative
opacities and adequate mixing length parameters (Sackmann \& Boothroyd 1991).
Yet, many recent AGB calculations
do not confirm those expectations (Vassiliadis \& Wood 1993, Wagenhuber
\& Weiss 1994, Bl\"ocker 1995, Forestini \& Charbonnel 1997), while some others
do succeed in obtaining dredge-up (Frost \& Lattanzio 1996, Straniero et al. 1997,
Herwig et al. 1997, this work).

  The disagreement between AGB model predictions on the issue of dredge-up
prevents us to provide
reliable and consistent chemical yields of low- and intermediate-mass stars.
Why do some models obtain dredge-up while others fail to do so?
Herwig et al. (1997) succeed in obtaining dredge-up
from the $7^{\mathrm{th}}$ pulse on in a \mass{3} Pop.~I star model
by using an overshooting algorithm below the convective envelope.
Some sort of extra-mixing beyond the convection borders defined by the
Schwarzschild criterion was already
used by Boothroyd \& Sackmann (1988a) and Lattanzio (1986), and
its inclusion is found to improve the dredge-up efficiencies predicted by the
models of Frost \& Lattanzio (1996).
These last authors further note the critical dependency of dredge-up on the
numerical treatment of convection in massive AGB models.
In contrast,
Straniero et al. (1997) claim to obtain dredge-up consistently {\it
without} invoking any extra-mixing. A similar claim is also reported by
Lattanzio (1989). According to Straniero et al. (1997), the solution resides
rather in the time and mass resolution of the models. Clearly,
several questions must be clarified. How influential is the numerical accuracy of
AGB models in obtaining dredge-up? What is the role of
extra-mixing below the envelope? Is extra-mixing {\it necessary} for
obtaining 3DUP in current AGB models using the Schwarzschild criterion? If yes,
how {\it sensitive} are the dredge-up characteristics to the extra-mixing parameters, about
which little is known today? And, finally, what {\it are} the characteristics
governing the 3DUPs?

The aim of this paper is to shed some light on these questions. They are, for a
great part, related to the delicate question of convection and the determination
of their boundaries in AGB models. The use of a local theory of convection, such as that of the
mixing length theory (MLT), is certainly one of the greatest shortcomings
still affecting stellar model calculations in general, and AGB models in
particular. New non-local formulations are being developed
(e.g. Canuto \& Dubovikov 1998) and may, hopefully, be included in future calculations. For the present
time, however, evolutionary AGB models are still performed using the MLT
prescription -- mainly due to computer time requirements --, and we also adopt it
in this study.
During the third dredge-up, however, the use of such a local prescription leads
to the development of an (unphysical) discontinuity in the abundance profiles when
the H-rich envelope penetrates into the H-depleted layers. The determination of
the lower boundary of the convective envelope then becomes problematic, and requires a
careful analysis of its stability against mixing across the discontinuity
(i.e. extra-mixing). That problematic is set forth in Sect.~\ref{Sect:boundaries}.
Several definitions involving the use of the Schwarzschild criterion and of extra-mixing
are also presented in that section. Sections~\ref{Sect:code} to \ref{Sect:dredge-up
laws} then analyze the 3DUP predictions from AGB model calculations of a \mass{3}
star of solar metallicity, computed both with and without extra-mixing.
Section~\ref{Sect:code} describes the main characteristics of the code used to
compute the models. A first set of calculations performed without any extra-mixing
is analyzed in Sect.~\ref{Sect:no overshooting}.
It is shown that dredge-up does not occur in those conditions.
In contrast, Sect.~\ref{Sect:overshooting} shows that models calculated with
extra-mixing lead naturally to the occurrence of the 3DUP phenomenon. The dredge-up
laws derived from those calculations are presented
in Sect.~\ref{Sect:dredge-up laws}. Finally, some implications
of those laws on stellar evolution along the AGB are analyzed
in Sect.~\ref{Sect:AGB evolution}.
Conclusions are drawn in Sect.~\ref{Sect:conclusions}.

\section{On the convective boundaries in AGB stars}
\label{Sect:boundaries}

  The criterion used in most stellar evolution calculations -- and in all AGB
models published to date -- to delimit convective regions is that of
Schwarzschild. According to that criterion, a layer is convective if
$\nabla_{rad}>\nabla_{ad}$, and radiative if $\nabla_{rad}<\nabla_{ad}$.
$\nabla_{ad}$ is the adiabatic temperature gradient, and $\nabla_{rad}$ the
radiative temperature gradient given by
\begin{equation}
\label{Eq:nabla_rad}
  \nabla_{rad} \equiv \left. \frac{{\rm d}\ln T}{{\rm d}\ln P} \right|_{rad}
               = \frac{3}{16 \pi a c} \frac{\kappa L P}{G m T^4}
\end{equation}
with $T$, $L$, $\kappa$, and $m$ being, respectively, the temperature,
luminosity, Rosseland mean opacity, and
mass contained in the sphere interior to the radius $r$ (other symbols having
their usual meanings).
The layer where $\nabla_{rad}=\nabla_{ad}$
defines the convection boundary, and is called the {\it Schwarzschild layer}.

Let us suppose for now that mixing of material inside a star can only occur
through convection, and let us imagine the star as being composed of a
finite number of shells located at increasing mass coordinates $m$. Each shell
is characterized by a temperature $T(m,t)$, density $\rho(m,t)$, luminosity $L(m,t)$
and chemical composition $X(m,t)$. All these quantities vary with time $t$. The
Schwarzschild criterion tests {\it locally} the stability of each shell
against convection by comparing $\nabla_{rad}(m,t)$ derived from
Eq.~\ref{Eq:nabla_rad} with $\nabla_{ad}$ at the given shell {\it independently} of the
values of those quantities in adjacent shells. Mixing of matter between two
adjacent shells is allowed to operate only if {\it both} of them are found to be convective.
Alternatively, the chemical composition of a radiative shell is not altered with
time, according to the Schwarzschild criterion, unless and until its radiative
gradient exceeds its adiabatic gradient. This could result from a change with
time of $T$,
$\rho$ and/or $L$ (we neglect here changes due to nuclear burning). Applied in
this strict definition, the local Schwarzschild criterion is called hereafter the
{\it strict Schwarzschild criterion}. We note that the shell-representation of a
star used in the above discussion corresponds to the numerical zoning
performed during the computation of stellar models (by the finite difference
method). Thus, stellar model calculations using the Schwarzschild criterion
without any extra-mixing essentially use the strict Schwarzschild criterion. For
this reason, the `strict Schwarzschild criterion' is equivalently called,
hereafter, the {\it Schwarzschild criterion without extra-mixing}.

  Convection, however, is essentially non-local. Convective bubbles reaching the
convection boundary {\it do} penetrate, on a whatever small distance, into the
radiative layers. The chemical composition of a radiative layer adjacent to a
convective zone may thus, in reality, be modified by this convective
penetration. For it to occur in models using the Schwarzschild criterion,
however, an extra-mixing procedure {\it must} be specified in the code (whether
purely numerically or according to a physical prescription). In most
cases, this extra-mixing does not affect the stability of the radiative layers
adjacent to the convective zone. A short discussion of these cases is presented in
Sect.~\ref{Sect:stable boundary}. In some cases, however, this extra-mixing {\it
does modify} the stability of those radiative layers. Those are discussed in
Sect.~\ref{Sect:unstable boundary}. They characterize, in particular, the
dredge-up phase in AGB stars. A third situation may be encountered in model stars
where both of the above cases are present simultaneously. This is discussed in
Sect.~\ref{Sect:minimim in gradrad}.

\subsection{Case a: Stable Schwarzschild boundary}
\label{Sect:stable boundary}

\begin{figure}
  \resizebox{\hsize}{!}{\includegraphics{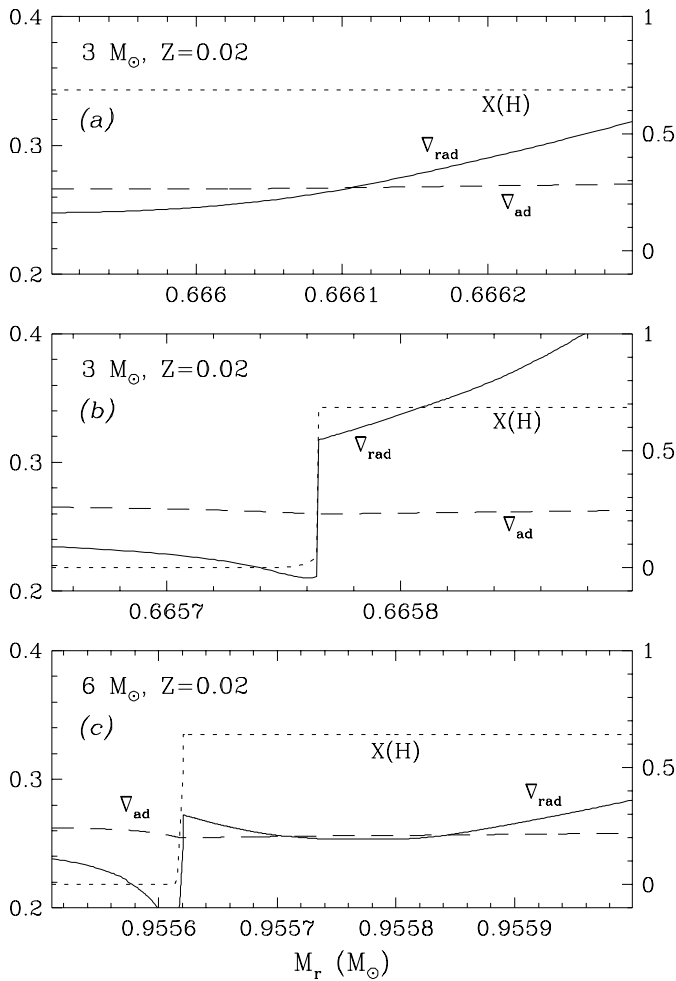}}
  \caption{Hydrogen mass fraction (dotted lines), adiabatic temperature gradient
           (dashed lines) and radiative temperature gradient (solid
           lines) profiles at the bottom of the convective envelope
           in three different AGB stars as a function of
           their mass coordinate:
           {\sl (a)} during the shell H-burning phase of a \mass{3} star;
           {\sl (b)} during the afterpulse phase of a \mass{3} star;
           {\sl (c)} during the afterpulse phase of a \mass{6} star.
          }
  \label{Fig:gradrad}
\end{figure}

 The most common case is encountered when the convection boundary lies
in a region of smooth $\nabla_{rad}$ profile, such as in
chemically homogeneous regions as illustrated in Fig.~\ref{Fig:gradrad}{\sl (a)}. This
case characterizes, in particular, the bottom layers of the convective envelope
in AGB stars during most of their pulse and interpulse phases.

  It is easily seen that an extra-mixing of material from the convective envelope
into the radiative layers does not alter the stability of those layers against
convection. The Schwarzschild layer is thus said to be stable in the sense
that an extra-mixing of matter from the convective zone into the underlying radiative
layers does not, to first order, alter the location of the Schwarzschild
layer.

  Of course, the Schwarzschild layer does not necessarily locate correctly the
boundary of the convective zone, since a penetration of the convective bubbles
into the radiative layers (i.e. overshooting) may occur, the extent of which is
still a question of debate.
But such an overshooting does not, to first approximation,  have any
crucial consequence on the location of the Schwarzschild layer.

\subsection{Case b: Unstable Schwarzschild boundary}
\label{Sect:unstable boundary}

During a 3DUP, the H-rich convective envelope penetrates into the H-depleted
radiative layers, and gives rise to a discontinuous chemical profile
[dotted line in Fig.~\ref{Fig:gradrad}{\sl (b)}]. As a result, a concomitant discontinuity
develops in the $\nabla_{rad}$ profile [solid line in Fig.~\ref{Fig:gradrad}{\sl
(b)}; see also Iben 1976, Paczy\'nski 1977, Frost \& Lattanzio 1996].
A similar situation occurs during the core
helium burning phase of most stars when the He-depleted convective core grows into the
overlying He-rich layers.

In models using the Schwarzschild criterion without extra-mixing, the border of
the convective zone is automatically located at the position of the hydrogen
abundance discontinuity. As a result, no alteration of any chemical abundance
is expected to occur in the radiative layers. The
convection boundary remains located at this position as long as the value of
$\nabla_{rad}$ on the radiative side of the discontinuity is
below $\nabla_{ad}$.
Evolutionary AGB model calculations performed without extra-mixing
confirm that no penetration of the convective envelope occurs beyond that point of
discontinuity (see Sect.~\ref{Sect:no overshooting}). It should be stressed that
the Schwarzschild layer is {\it undefined} in those models, a situation which is
clearly unphysical.

  In reality, some matter does travel across the $\nabla_{rad}$
discontinuity. This results from the finite velocity of the bubbles at the
convection boundary, since $\nabla_{rad}-\nabla_{ad}$ is positive.
Models should thus include some sort of extra-mixing. The
consequences of such a penetration can readily be evaluated:
mixing of matter from the
envelope into the radiative regions enriches the latter region with hydrogen
and renders them unstable against convection. This in turn leads to further penetration of
the envelope. The situation is thus {\it unstable}, in the sense that an
extra-mixing of matter from the convective zone into the underlying layers does
alter the location of the convective boundary.

The challenge facing AGB modelers is to provide a reliable description of
this envelope penetration with simple (i.e. computer feasible) convection prescription.
The analysis presented in Sect.~\ref{Sect:overshooting} actually reveals that
the penetration of the envelope into the H-depleted layers is
rather insensitive to the extra-mixing parameters such as its
extent or mixing efficiency. That main conclusion to be obtained in
Sect.~\ref{Sect:overshooting} is most welcomed since not much -- not to say
nothing -- is known on the extra-mixing characteristics in AGB stars.
Section~\ref{Sect:overshooting} also shows that the
use of a diffusive algorithm for the mixing of chemical elements enables to avoid
the development of a $\nabla_{rad}$ discontinuity in the models, and to properly
define the Schwarzschild layer.

\subsection{Case c: Existence of a minimum in the $\nabla_{rad}$ profile}
\label{Sect:minimim in gradrad}

In massive AGB stars, the $\nabla_{rad}$ profile can display a minimum close to the
lower boundary of the envelope, as illustrated in Fig.\ref{Fig:gradrad}{\sl (c)}.
A situation can occur where two convective zones develop, separated by a radiative
region around that minimum. 
This case combines the features of both previous cases: the outer convective zone
has a smooth $\nabla_{rad}$ profile as in case {\sl a}, while the inner one
presents a discontinuity as in case {\sl b}.

In models using the Schwarzschild criterion with no extra-mixing, the inner
convective zone does
not penetrate into the H-depleted regions due to the $\nabla_{rad}$
discontinuity (see Sect.~\ref{Sect:unstable boundary}). The outer zone, on the
other hand, is free to deepen until it
reaches the inner convective zone. The two convective zones then merge
together, and the situation resumes to case {\sl b}.

When extra-mixing is allowed to operate, a specific scheme is expected to develop.
Because of the small amount of mass involved in the inner convective zone
($\sim\mmass{10^{-4}}$), any extra-mixing
into the He-rich layers reduces the hydrogen content of that convective zone,
and renders its outermost
layers stable. Those outermost layers are then progressively excluded
from the convective zone as they reach convective neutrality, and a
semi-convective zone is expected to develop\footnote{A semi-convective zone is defined here to
be one throughout which partial mixing leads to convective neutrality
$\nabla_{rad}=\nabla_{ad}$ (see, e.g., Mowlavi \& Forestini 1994). This definition is
not to be confused with other ones used in the literature, such as that
whereby a semi-convective region is defined by
$\nabla_{ad}\le\nabla_{rad}\le\nabla_{Led}$, $\nabla_{Led}$ being the Ledoux
adiabatic gradient (which takes into account the gradient of the mean molecular
weight). Furthermore, the semi-convective zone discussed here, and
located at the H-He discontinuity,
is different from that described by Iben (1981), which is located at the He-C
discontinuity.}. Eventually, the main convective envelope penetrates into
that semi-convective zone, and the case resumes to case {\sl b}.

\subsection{Conclusions}

It is clear from the discussion in this section that models using the
Schwarzschild criterion with no extra-mixing are inadequate to describe the 3DUP
phenomenon. There is, however, a great deal of confusion in the literature
on the third dredge-up predictions, and in particular on the role of
extra-mixing. For the purpose of clarifying some of those issues, AGB models have
been computed both with and without extra-mixing. Particular attention
has been devoted to ensure that the code comply with the definitions given at the
beginning of this section on the location of convection boundaries. The description and
results of those calculations are presented in Sects.~\ref{Sect:code} to
\ref{Sect:overshooting}.

\section{The stellar evolution code}
\label{Sect:code}

The stellar evolution code used to compute the models is described in detail
in Mowlavi (1995). Among the modifications brought to it since its conception
in 1995, let me
mention the use of interior radiative opacities taken from Iglesias \& Rogers (1996)
and the use of low-temperature ones from Alexander \& Ferguson (1994). The
convection energy transport is described as usually by the MLT, with the
mixing length parameter $\alpha_{mlt}$ taken equal to 1.5 in units of pressure
scale height. The time-step and mesh allocation procedures relevant to our study
are presented in more details in Sect.~\ref{Sect:model accuracy}, while
the mixing schemes of chemical elements are described in
Sect.~\ref{Sect:mixing}.

\subsection{Time-step and mesh distribution}
\label{Sect:model accuracy}

The time step between each model is determined, in most cases, by the relative
variation of the dependent variables (radius, temperature, pressure and luminosity)
from one model to the next such that it does not exceed 10\% at any layer in the
star. On the AGB phase, the time step is further limited to a maximum of
$\Delta t_{interp}/1000$, where $\Delta t_{interp}$ is the time between
two successive pulses. The time-step during the 19th interpulse phase,
for example, is of the order of 25~y, and can decrease down to few hours
during a pulse.

The mesh allocation is based on the constitutive differential equations,
and ensures a relative accuracy in the finite difference
method\footnote{Our criterion for the mesh allocation
contrasts with that used in almost all
existing stellar evolution codes where the mesh distribution is determined
according to
the relative variation of the dependent variables (except that of
Wagenhuber \& Weiss 1994 who use a method similar to ours).} of better than
$5\times 10^{-5}$ (see Mowlavi 1995).
Special care is taken around boundaries of
convective regions and in regions with steep chemical
abundance profiles in order to avoid spurious numerical chemical diffusion.
Our models are characterized by about 3000 meshes during the interpulse
phases, and by up to 8000 meshes during pulses.

\subsection{Mixing}
\label{Sect:mixing}

Instantaneous mixing is assumed in all convective zones in models with no
extra-mixing. 
When a discontinuity in the $\nabla_{rad}$ profile is detected at the convection
border, care is taken {\it not} to mix the material of the mesh comprising that
discontinuity with the convective zone. This prevents a spurious numerical
propagation of the convective zone into the radiative layers in unstable
cases such as case {\sl b} described in Sect.\ref{Sect:unstable boundary}, and
guarantees a correct application of the Schwarzschild criterion described in
Sect.~\ref{Sect:boundaries}.

In models with extra-mixing, a specific overshooting prescription is used beyond the
Schwarzschild layer. The bubble velocity field $v(r)$ in the radiative
field is assumed to decay exponentially with the distance to the Schwarzschild
layer, in the form
\begin{equation}
\label{Eq:v(r)}
  v(r) = v_{sch} \times x(r) \times
         \exp\left[ \ln(f_v)\times\frac{(r_{sch}-r)}{(r_{sch}-r_{ov})} \right],
\end{equation}
with
\begin{equation}
\label{Eq:f_v}
  f_v=\frac{v_{ov}}{v_{sch}},
\end{equation}
where subscripts $sch$ and $ov$ refer to quantities evaluated at the
Schwarzschild boundary $r_{sch}$ and at the edge of the overshooting region
$r_{ov}$, respectively. The function
$x(r)=[(r-r_{ov})/(0.5\, r+0.5\, r_{sch}-r_{ov})]^{0.1}$ ensures that $v(r)$ vanishes
at the outer edge of the overshooting region. It is close to unity in
all the overshooting region except close to $r_{ov}$ where it vanishes.
Mixing in the convective $+$ overshooting
regions is performed through an algorithm coupling
diffusive mixing and nucleosynthesis (see Mowlavi 1995)
with a diffusion coefficient equal to $D_{conv}(r)=\frac{1}{3}\times v(r)\times
l_{mix}(r)$, where $l_{mix}(r)=\alpha_{mlt}\times H_p(r)$ ($H_p$ being the
pressure scale height).
The bubble speed at $r_{sch}$, $v_{sch}$, is determined from the bubble velocity
profile in the convective zone as given by the MLT, while that at $r_{ov}$, $v_{ov}$, is
determined by an `efficiency' parameter $f_v$ according to Eq.~\ref{Eq:f_v}.
The extent of the overshooting region $(r_{sch}-r_{ov})$ is determined
by a second parameter, $\alpha_{ov}$, defined such that $r_{ov}$ locates
the layer where $P_{ov}=P_{sch}\times \exp(\alpha_{ov})$, $P_{ov}$ and $P_{sch}$
being the pressure at $r_{ov}$ and $r_{sch}$, respectively.

In those models with extra-mixing, the mesh distribution is increased
in the overshooting regions by about 100 meshes in order to ensure a good
description of the abundance profiles. The time-step is further reduced to about
$5\times 10^{-2}$~y during a 3DUP. A dredge-up episode is then covered by about
2000-3000 models.

Let us remark that a velocity field of the type of Eq.~\ref{Eq:v(r)}
is supported by recent hydrodynamical simulations of stellar convection
in solar-type stars and white dwarfs (Freytag et al. 1996).
However, we do not claim that such an overshooting pattern also applies
to the envelopes of AGB stars, in which the physical conditions are quite
different from those characterizing solar-type stars or white dwarfs.
That remains to be confirmed by future hydrodynamical studies.
We rather use such a prescription merely as a way to simulate an extra-mixing
below the envelope and to analyze its effects on the 3DUP predictions.
The extra-mixing could
very well be associated with other `non-standard' mixing processes such
as, for example, diffusion induced by rotation or shear mixing. Actually, the results
presented in Sect.~\ref{Sect:overshooting} show that the dredge-up
predictions are quite insensitive to the extra-mixing parameters.

\section{Models without extra-mixing}
\label{Sect:no overshooting}

\subsection{Failure to obtain dredge-up}

\begin{figure}
  \resizebox{\hsize}{!}{\includegraphics{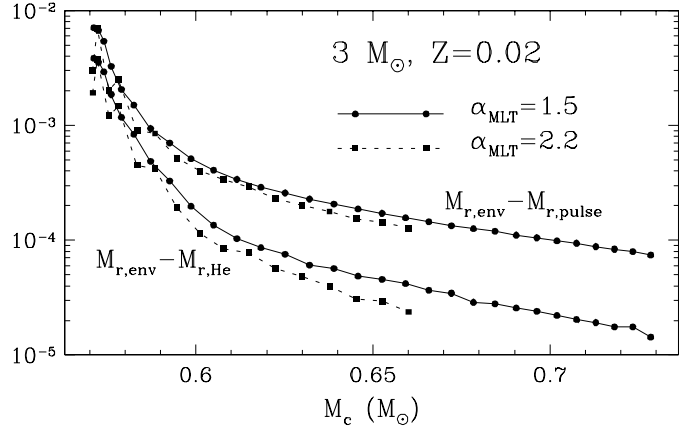}}
  \caption{Depth reached, as a function of the core mass, by the envelope of
           \mass{3}, $Z$=0.02 models (where $Z$ is the metallicity) during the
           first 20 to 30 thermal pulses (identified by the filled circles and
           squares on each curve). The two lower curves, labeled
           $M_{r,env}-M_{r,He}$, indicate the mass, in solar mass units, separating the
           lower boundary of the envelope to the location of the H-He discontinuity. The two
           upper curves give a similar information, but to the location of the He-C
           discontinuity. Solid and dashed lines refer to models calculated with
           $\alpha_{MLT}=1.5$ and 2.2, respectively.
          }
  \label{Fig:3noov}
\end{figure}

A set of standard\footnote{The word `standard'
refers throughout this paper to models calculated without any extra-mixing beyond the
Schwarzschild layer.}
\mass{3} models of solar metallicity is computed with $\alpha_{mlt}=1.5$
from the pre-main sequence phase up to the 32nd pulse on the TP-AGB phase.
The maximum depth reached by the convective envelope after each pulse
is shown in Fig.~\ref{Fig:3noov}. The depth is expressed in terms
of the mass separating the lower convective envelope boundary to either the H-He
discontinuity (lower solid line) or the He-C one (upper solid lines). It is seen that
{\it the envelope does not penetrate into the He-rich layers} (and a fortiori into
the C-rich layers) during the first thirty two pulses.

\begin{figure}
  \resizebox{\hsize}{!}{\includegraphics{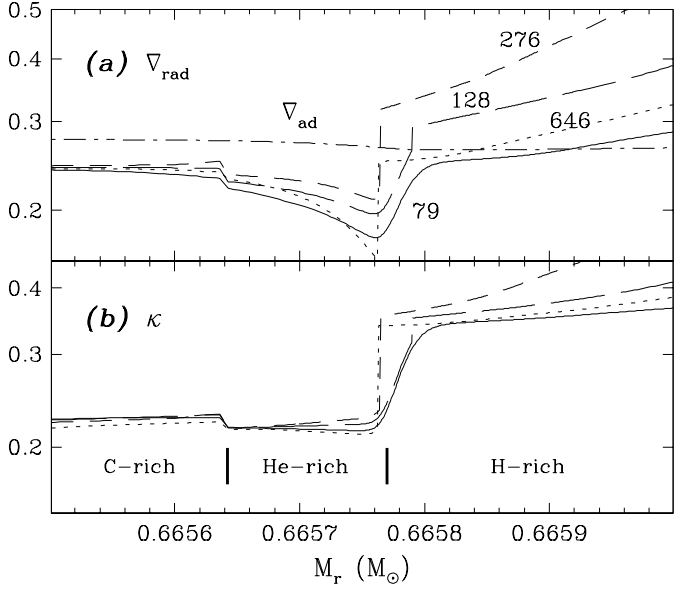}}
  \caption{{\it(a)} Radiative gradient profile at several
           times, as labeled (in yr) on the curves, after the 19th pulse of
           the standard \mass{3} case. The time origin is
           arbitrary but identical to that of Fig.~\ref{Fig:pulse19}. The dashed
           dotted line is the adiabatic gradient taken at the time
           corresponding to the solid line; {\it(b)} Same as {\it(a)}, but for
           the opacity profile. The extents of the H-, He- and C-rich regions
           are shown at the bottom of {\it(b)}, separated by vertical lines}
  \label{Fig:gradnoov}
\end{figure}

The non-occurrence of the third dredge-up in those models is easily understood as
resulting from the discontinuity in the H abundance when the convective envelope
reaches the H-He discontinuity (case {\sl b} described in Sect.~\ref{Sect:unstable
boundary}).
In those regions, the opacity is mainly determined by the Thompson opacity
$\kappa_{th}=0.2(1+X)$, where $X$ is the hydrogen mass fraction.
It presents a jump
across the composition discontinuity, with the H-rich layers being more opaque to
radiation than the He-rich layers [Fig.~\ref{Fig:gradnoov}{\it(b)}].
This discontinuity in $\kappa$ translates into a similar discontinuity
in $\nabla_{rad}$ [Fig.~\ref{Fig:gradnoov}{\it(a)}], which prevents
further penetration of the envelope into
the H-depleted regions {\it when no extra-mixing is allowed in the models}.

\begin{figure}
  \resizebox{\hsize}{!}{\includegraphics{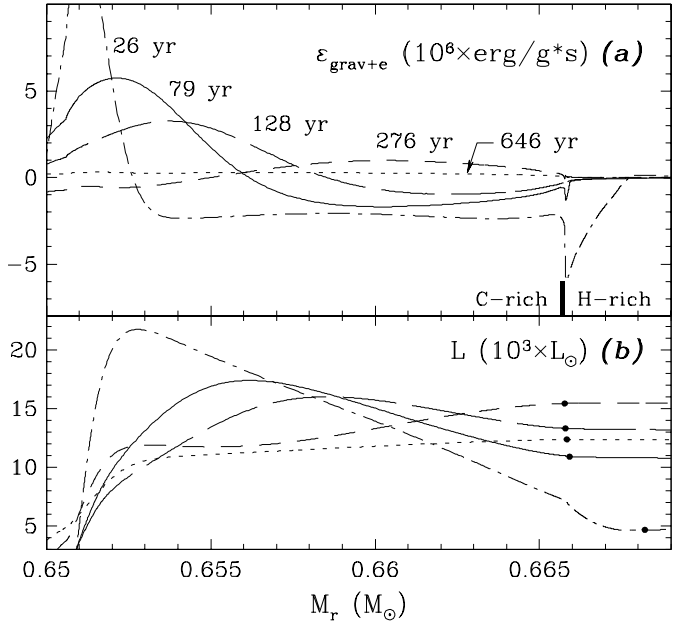}}
  \caption{{\it(a)} Gravothermal energy profiles at 5 different times, as labeled
           (in yr) on
           the curves, after the 19th pulse of the standard \mass{3}, solar metallicity
           star. The time origin is taken at maximum pulse extension.
           The He-rich region is identified by the thick
           vertical line separating the H- and C-rich regions;
           {\it(b)} Same as {\it(a)}, but for the luminosity profiles. The thick
           dots on the curves indicate the location of the lower convective envelope
           boundary}
  \label{Fig:epsgrav}
\end{figure}

It is instructive to recall
that the structural readjustments characterizing the afterpulse phases result
from the evacuation of a gravothermal energy (defined by
$\epsilon_{grav+e} = P\; \frac{{\rm d}(1/\rho)}{{\rm d}t} + \frac{{\rm d}e}{{\rm
d}t}$,
where $P$, $\rho$ and $e$ are the pressure, density and internal energy,
respectively) which develops at the bottom of the
former pulse (Paczy\'nski 1977). This translates into a positive gravothermal
energy wave propagating
outwards and reaching the H-rich layers in about 250 yr during the 19th
afterpulse in our \mass{3} models (see Fig.~\ref{Fig:epsgrav}).
The luminosity at the bottom of the envelope, $L_{env,b}$, increases concomitantly
with time until the gravothermal energy has been evacuated
[Fig.~\ref{Fig:epsgrav}{\it(b)}]. As a result,
$\nabla_{rad}$ increases too, and the convective envelope penetrates inwards
until it reaches the H-He discontinuity.
A further deepening of the envelope could be possible if the
increase in $L_{env,b}$ overcomes the discontinuity in $\nabla_{rad}$. In the
standard \mass{3} models, however, this does not happen during the
first thirty two pulses calculated so far.

\paragraph{Effect of mixing length parameter}

The mixing length parameter is known to be very influential on the depth of
convective envelopes (Wood 1981). In order to analyze its role in obtaining 3DUP
in AGB models without extra-mixing, the standard \mass{3} star is recalculated
from the first to the 23rd pulse with $\alpha_{MLT}=2.2$ (i.e. similar to the
value used by Straniero et al. 1997). The results are shown in dotted lines in
Fig.~\ref{Fig:3noov}. Indeed the convective envelope reaches deeper layers with
$\alpha_{MLT}=2.2$ than with 1.5. But they are not more successful in penetrating the
He-rich layers. This contrasts with the results of Straniero et al.
who claim to obtain dredge-up in their models without any extra-mixing.

\paragraph{Effect of stellar mass}

\begin{figure}
  \resizebox{\hsize}{!}{\includegraphics{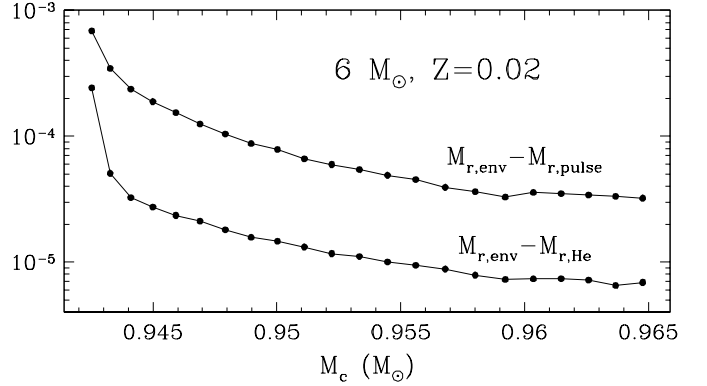}}
  \caption{Same as Fig.~\ref{Fig:3noov}, but for \mass{6}, Z=0.02 models
with $\alpha_{MLT}=1.5$}
  \label{Fig:6noov}
\end{figure}

Finally, a last set of standard models is calculated for a \mass{6} star of solar
metallicity up to its 22nd pulse. It is indeed known that 3DUP should also be
favored in more massive AGB stars (Wood 1981). The results
of our model calculations are shown in Fig.~\ref{Fig:6noov}. Again, {\it no}
dredge-up is found in those models calculated without any extra-mixing, for
reasons similar to those put forward for the \mass{3} standard models.

\paragraph{Effect of numerical accuracy}

  The analysis presented in this section shows that models using the local
Schwarzschild criterion without extra-mixing do not lead to the occurrence
of third dredge-up because of the $\nabla_{rad}$ discontinuity at the core's edge.
That conclusion should thus not be sensitive to the numerical accuracy of the
models.
Indeed,
test calculations performed on the 19th afterpulse without
extra-mixing, but with increased or decreased accuracies on both the
time-step and mesh resolution lead to results identical to those
presented in this section.

\subsection{Useful relations}
\label{Sect:relations}

  Some relations characterizing the evolution of the standard \mass{3}
models are presented in this section. These will be useful in
Sect.~\ref{Sect:dredge-up laws}.

\begin{figure}
  \resizebox{\hsize}{!}{\includegraphics{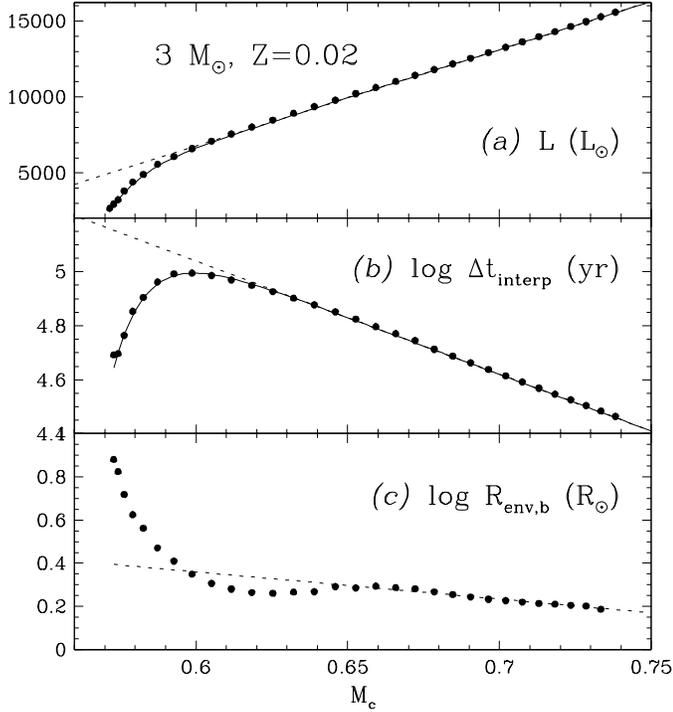}}
  \caption{{\sl (a)} Stellar luminosities at maximum pulse extensions,
           {\sl (b)} interpulse pulse duration between two maximum
           pulse extensions and {\sl (c)} location of the lower envelope
           boundary at its first minimum after each pulse in the standard
           \mass{3} model calculations (filled circles in each panel).
           The dashed lines show the linear relations
           fitting the data in each panel (Eqs.~\ref{Eq:McL linear},
           \ref{Eq:Mcdt linear} and \ref{Eq:McenvRb linear}).
           The solid lines in panels {\sl (a)} and {\sl (b)} show
           the $M_c-L$ (Eq.~\ref{Eq:McL}) and $M_c-\Delta t_{interp}$ (Eq.~\ref{Eq:Mcdt})
           relations, respectively, when the deviation from linearity of the
           first pulses is taken into account.
          }
  \label{Fig:relations}
\end{figure}

  The first relation describes the evolution of the surface luminosity $L$ with
core mass growth. The values of those quantities before the onset of each pulse
of our standard \mass{3} models are displayed as filled circles in
Fig.~\ref{Fig:relations}{\sl(a)}. It is well known that AGB models not
experiencing dredge-up reach an asymptotic regime characterized by a linear
relation between $L$ and $M_c$. This is the famous $M_c-L$ relation, which writes
from our \mass{3} models [dotted line in Fig.~\ref{Fig:relations}{\sl(a)}]
\begin{equation}
\label{Eq:McL linear}
  L= 63340\; (M_c - 0.493),
\end{equation}
$L$ and $M_c$ being given in solar units.
A correction has to be applied to this relation
in order to account for the lower luminosities of the first pulses. This
correction is found to be well reproduced by an exponential function, and
Eq.~\ref{Eq:McL linear} becomes
\begin{equation}
  \label{Eq:McL}
  L= 63340\; (M_c - 0.493) - 1100 \;\times\; \mathrm{e}^{-(100\, M_c + 58)}.
\end{equation}
This relation is shown in solid line in Fig.~\ref{Fig:relations}{\sl(a)}, and is
seen to fit very well model predictions.

  The second useful relation expresses the interpulse duration $\Delta t_{inter}$
as a function of core mass growth. From our standard \mass{3} model calculations,
shown in filled circles in Fig.~\ref{Fig:relations}{\sl(b)},
we find an asymptotic $M_c-\log \Delta t_{interp}$ relation given by [dotted line
in Fig.~\ref{Fig:relations}{\sl(b)}]
\begin{equation}
\label{Eq:Mcdt linear}
  \log \Delta t_{interp} = 4.2\; (1.8-M_c),
\end{equation}
$\Delta t_{interp}$ being expressed in yr.  Again, a correction
has to be applied during the first pulses. The resulting relation writes [solid
line in Fig.~\ref{Fig:relations}{\sl(b)}]
\begin{equation}
\label{Eq:Mcdt}
  \log \Delta t_{interp} = 4.2\; (1.8-M_c) - 0.1 \;\times\; \mathrm{e}^{-(90\, M_c-53.2)}.
\end{equation}

  Finally, the location $R_{env,b}$ of the lower envelope boundary at its first
minimum after the pulse is displayed in filled circles in
Fig.~\ref{Fig:relations}{\sl(c)} as a function of $M_c$. A linear relation fits
the data in the asymptotic regime [dotted line in
Fig.~\ref{Fig:relations}{\sl(c)}], which writes in solar units
\begin{equation}
\label{Eq:McenvRb linear}
  R_{env,b} = 1.263 \times (0.885-M_c).
\end{equation}

\section{Dredge-up in models with extra-mixing}
\label{Sect:overshooting}

\begin{figure}
  \resizebox{\hsize}{!}{\includegraphics{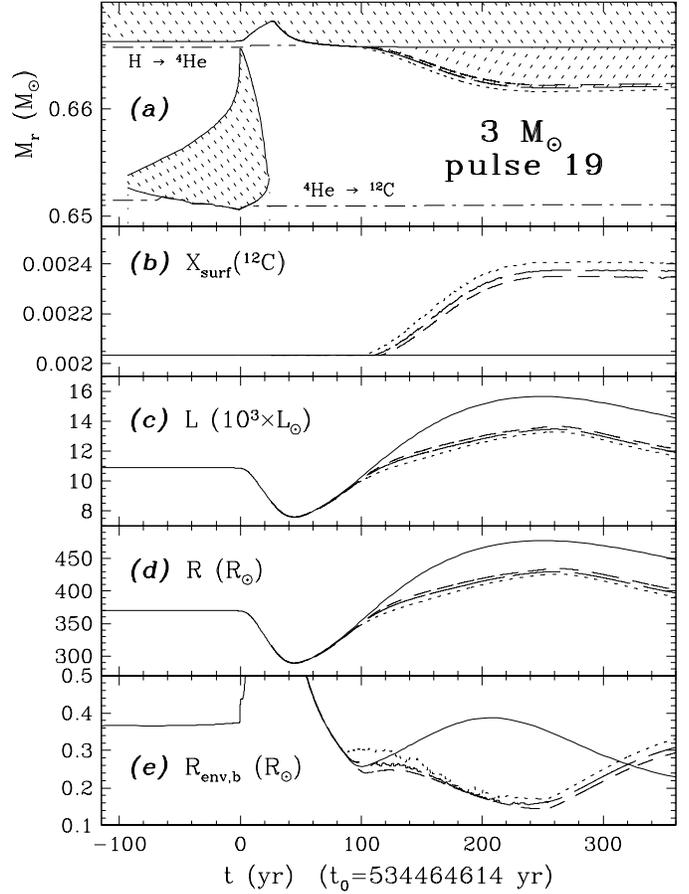}}
  \caption{Structural evolution of several quantities during and after the 19th pulse of the
           \mass{3} star in the standard case (solid line)
           and in models with envelope overshooting
           (dotted line: $\alpha_{ov}=0.20, f_v=10^{-6}$;
            long dashed line: $\alpha_{ov}=0.60, f_v=10^{-4}$;
            short dashed line: $\alpha_{ov}=1.20, f_v=10^{-3}$).
           {\it(a)} Mass location of the convective boundaries and of the
           layer of maximum nuclear energy
           production (dashed-dotted lines) in the H and He burning shells, as
           labeled in the graph. Hatched regions correspond to convective
           zones. The extent of the overshooting regions is less than
           \mass{10^{-4}} and cannot be distinguished in the figure;
           {\it (b)} surface \chem{C}{12} mass fraction;
           {\it (c)} stellar luminosity;
           {\it (d)} stellar radius;
           {\it (e)} location of the lower boundary of the envelope.
           The time origin is taken at maximum pulse extension.}
  \label{Fig:pulse19}
\end{figure}

\begin{figure}
  \resizebox{\hsize}{!}{\includegraphics{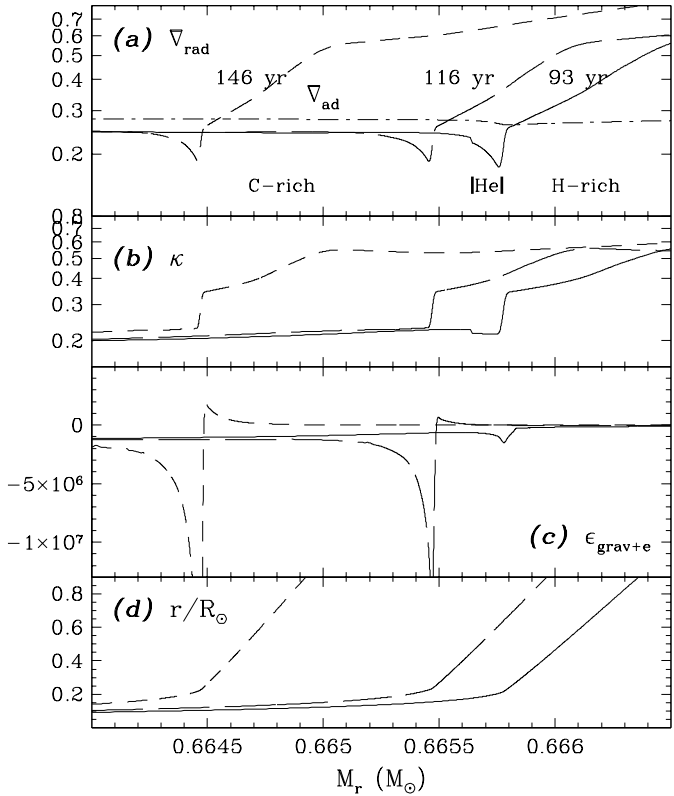}}
  \caption{{\it(a)} Same as Fig.~\ref{Fig:gradnoov}{\it(a)},
           but for models with overshooting ($\alpha_{ov}=0.60$,
           $f_v=10^{-4}$).
           {\it(b)}, {\it(c)} and {\it(d)} same as {\it(a)}, but
           for the opacity, the gravothermal energy and the radius,
           respectively.
          }
  \label{Fig:gradov}
\end{figure}

\subsection{Model results}
\label{Sect:dredge-up calculations}

The structural evolution of the 19th afterpulse of the standard \mass{3} star
is recalculated in a set of models using the overshooting prescription
described in Sect.~\ref{Sect:mixing}, with $(\alpha_{ov},f_{v})=(0.60,10^{-4})$.
The structural evolution of the intershell layers of these models is shown in
long dashed line in Fig.~\ref{Fig:pulse19}{\sl (a)}.

 Clearly, the use of extra-mixing
leads to the penetration of the convective envelope into the H-depleted
regions, as expected from the discussion in Sect.~\ref{Sect:unstable boundary}.
The role of overshooting in the 3DUP was already 
addressed by Iben (1976) and Paczy\'nski (1977). The latter author, in
particular, showed the essential role of extra-mixing in obtaining 3DUP.
The envelope is seen in our models to deepen at a maximum dredge-up rate of
\mass{4.2\times 10^{-5}}/yr. The dredge-up lasts for about 120~yr, which
corresponds to the time necessary for the gravothermal energy to
evacuate from the bottom of the formal pulse to the stellar surface
(see Fig.~\ref{Fig:epsgrav}). \mass{3.6\times 10^{-3}} of
C-rich material is mixed into the envelope, which leads to
a 16\% increase in the surface carbon mass fraction.

\begin{figure}
  \resizebox{\hsize}{!}{\includegraphics{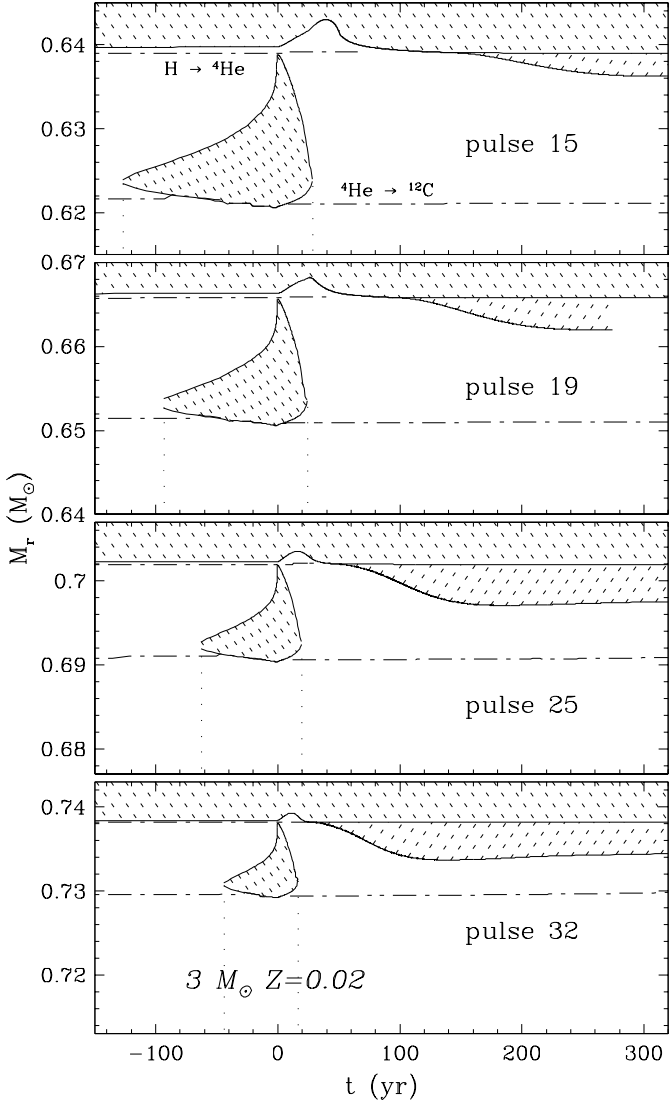}}
  \caption{Same as Fig.~\ref{Fig:pulse19}{\sl (a)}, but for, respectively from top to bottom,
           the 15th, 19th, 25th and 32nd pulses of the \mass{3} models
           with $\alpha_{ov}=0.60$ and $f_v=10^{-4}$. The size of the
           mass ordinate is similar in all four figures.
          }
  \label{Fig:pulse15_32}
\end{figure}

\begin{table*}
\caption[]{\label{Tab:dredge-ups}
           Several quantities characterizing the afterpulse phases of the 12th,
           15th, 19th, 25th and 32nd pulses of the \mass{3} star computed with the overshooting
           parameters $\alpha_{ov}=0.60$ and $f_v=10^{-4}$: the core mass $M_c$ and the
           luminosity $L$ at maximum pulse extension, the luminosity dip $L_{min}$
           during the afterpulse phase [see Fig.~\ref{Fig:pulse19}{\sl(c)}], the maximum
           dredge-up rate of the envelope $r_{dup}$, the core mass increase $\Delta
           M_{c,nuc}$ due to nuclear burning during the preceding interpulse phase, the mass
           of C-rich material dredged-up $M_{dup}$, the efficiency parameter $\lambda$, the
           carbon mass fraction $X_p(\mchem{C}{12})$ in the intershell layers left over by
           the pulse, and the increase in the envelope \chem{C}{12} mass fraction $\Delta
           X_s(\mchem{C}{12})$ resulting from each dredge-up.
          }
\begin{tabular}{c c r r c c c c c c}
\hline
\noalign{\smallskip}
 pulse  & $M_c$   &  $L\;\;\;$    & $L_{min}$ & $r_{dup}$ & $\Delta M_{c,nuc}$ & $M_{dup}$ & $\lambda$ & $X_p(\mchem{C}{12})$ & $\Delta X_s(\mchem{C}{12})$\\
 number &(\mass{})&(\lsun{})& (\lsun{}) &(\mass{}/y)&  (\mass{})  & (\mass{}) &           &                      & \\
\noalign{\smallskip}
\hline
\noalign{\smallskip}
 12  & 0.6184 &  7928 &  5015 & $1.45\;10^{-5}$ & 0.00665 & 0.00223 & 0.335 & 0.239 & 0.095 \\  
 15  & 0.6390 &  9260 &  6117 & $2.83\;10^{-5}$ & 0.00679 & 0.00354 & 0.521 & 0.222 & 0.247 \\  
 19  & 0.6657 & 10912 &  7600 & $4.25\;10^{-5}$ & 0.00650 & 0.00437 & 0.672 & 0.206 & 0.337 \\  
 25  & 0.7019 & 13140 &  9760 & $6.55\;10^{-5}$ & 0.00566 & 0.00524 & 0.926 & 0.200 & 0.566 \\  
 32  & 0.7382 & 15575 & 12720 & $7.34\;10^{-5}$ & 0.00480 & 0.00473 & 0.985 & 0.199 & 0.388 \\  
\noalign{\smallskip}
\hline
\end{tabular}
\end{table*}

\begin{figure}
  \resizebox{\hsize}{!}{\includegraphics{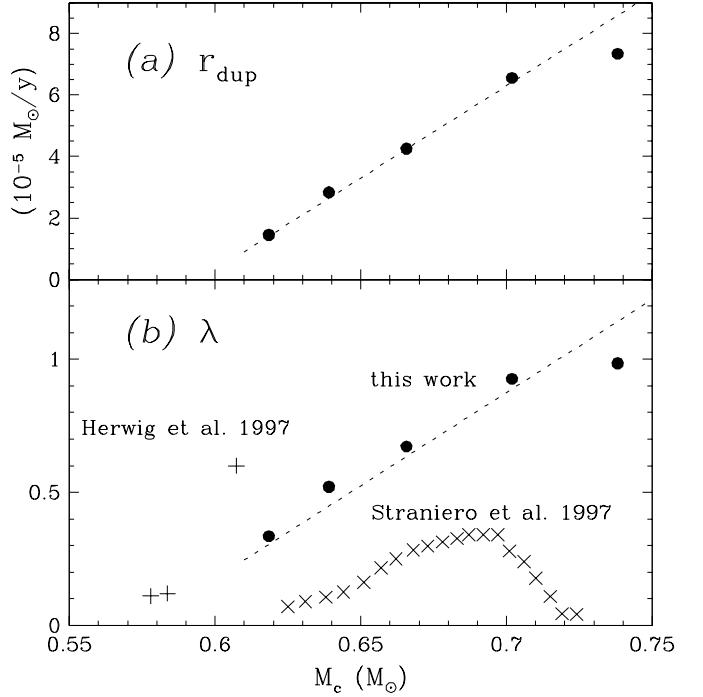}}
  \caption{{\sl (a)} Dredge-up rate and {\sl (b)} dredge-up efficiency
           as a function of core mass in the models
           with $\alpha_{ov}=0.60$ and $f_v=10^{-4}$ (filled dots).
           The dotted lines represent a linear fit to the data
           (Eqs.~\ref{Eq:r_dup} and \ref{Eq:lambda} for $v_{dup}$
           and $\lambda$, respectively). The crosses and plus signs in panel {\sl
           (b)} are the data predicted by Straniero et al. (1997) and Herwig et al. (1997),
           respectively.
          }
  \label{Fig:lambda}
\end{figure}

It is instructive to stress that the extent of the overshooting region below the envelope
comprises less than
\mass{10^{-4}}. This is 40 times less than the total amount of material dredged-up.
Overshooting alone can thus not explain the amount of material dredged-up.
Indeed, the occurrence of dredge-up essentially results
from the unstable character of the convection border in
the standard models (see Sect.~\ref{Sect:unstable boundary}).
Extra-mixing has the role of {\it triggering} the penetration of the envelope into
the H-rich layers. Had we, at any time during the dredge-up, suppressed the
extra-mixing procedure in the model calculations, the envelope would have stopped
from further penetration for reasons described in Sect.~\ref{Sect:no overshooting}.

The use of a {\it diffusive} extra-mixing procedure results in
models displaying smooth $\nabla_{rad}$ profiles as shown in Fig.~\ref{Fig:gradov}{\sl (a)}.
This enables to locate properly the Schwarzschild layer.
It should be noted, however, that the Schwarzschild layer
is still unstable in the sense that more mixing of hydrogen from the envelope into
the deeper layers renders them unstable against convection, leading thereby to further
penetration of the envelope.

The question then arises of
the sensitivity of the dredge-up characteristics to the extra-mixing parameters.
In order to answer that question, two other sets of models are
calculated, one with increased extra-mixing extent and efficiency of
$(\alpha_{ov},f_{v})=(1.20,10^{-3})$, and one with lower extra-mixing extent
and efficiency of $(\alpha_{ov},f_{v})=(0.20,10^{-6})$. The predictions from those
sets are displayed in Fig.~\ref{Fig:pulse19} in dotted and short-dashed lines,
respectively. It is seen that {\it the dredge-up characteristics are rather
insensitive to the extra-mixing parameters}.
In particular, the dredge-up rate is
independent of the extra-mixing parameters, and equals \mass{4.2\times 10^{-5}}/yr
in all three sets of calculations. The only small difference between the
different sets is the earlier time, in models with higher $\alpha_{ov}$, at which
the envelope reaches the H-depleted regions.

The important conclusion that the dredge-up characteristics are rather insensitive to the
extra-mixing parameters can be understood from basic considerations on the
dredge-up process, developed in Sect.~\ref{Sect:dredge-up laws}. It allows, in
particular, to analyze the 3DUP phenomenon without worrying much about the exact
values of the overshooting parameters. In order to perform such an analysis
as a function of stellar parameters, additional calculations with
extra-mixing are performed on the 12th, 15th, 25th and 32nd afterpulses
of the standard \mass{3} star. The resulting structural
evolution of the intershell layers are shown in
Fig.~\ref{Fig:pulse15_32}, and some of their characteristics summarized
in Table~\ref{Tab:dredge-ups}.

\paragraph{Dredge-up rate}

  The dredge-up rate $r_{dup}$ is seen from Table~\ref{Tab:dredge-ups}
to be an increasing function of the pulse number.
A linear relation between $r_{dup}$ and $M_c$ is
found from the model calculations, as shown in Fig.~\ref{Fig:lambda}{\sl
(a)}. It translates into the following equation for our \mass{3} star [dashed
line in Fig.~\ref{Fig:lambda}]
\begin{equation}
\label{Eq:r_dup}
  r_{dup} = 6\times 10^{-4}\; (M_c - 0.595) ,
\end{equation}
$M_c$ and $r_{dup}$ being expressed in \mass{} and \mass{} yr$^{-1}$,
respectively.

\paragraph{Dredge-up efficiency}

 The depth reached in mass coordinates by the envelope after each pulse
depends on both the time over which it penetrates and the dredge-up rate.
The former is mainly determined 
by the thermal time-scale for the evacuation of the gravothermal
energy accumulated at the bottom of the pulse (Fig.~\ref{Fig:epsgrav}), while the
latter is determined by the time-scale of the thermal readjustment of the envelope.
The resulting mass of C-rich
material dredged-up in the envelope, $M_{dup}$, is reported in
Table~\ref{Tab:dredge-ups}.

The dredge-up efficiency is traditionally expressed by a number $\lambda$, which is
the ratio of $M_{dup}$ to the core mass increase $\Delta M_{c,nuc}$
due to hydrogen burning during two successive pulses.
A value of $\lambda=1$, for example, implies a constant core mass from one pulse
to the next. From the model calculations, $\lambda$
is found to also increase linearly with $M_c$
[filled circles in Fig.~\ref{Fig:lambda}{\sl (b)}]. The linear relation writes [dotted
line in Fig.~\ref{Fig:lambda}{\sl (b)}]
\begin{equation}
\label{Eq:lambda}
  \lambda= 7\times\;(M_c - 0.571),
\end{equation}
$M_c$ being expressed in solar mass.

\paragraph{Evolution of surface characteristics during the dredge-up}
\label{Sect:surface characteristics}

 The evolution of the surface luminosity and radius during the 3DUP is
shown in Fig.~\ref{Fig:pulse19}{\sl (c,d)} (solid lines),
and compared to that predicted by models without dredge-up (dotted lines). It is
seen that dredge-up leads to surface luminosities lower than those obtained in
otherwise similar models but without dredge-up. That the luminosity should be lower in
models with dredge-up is easily understood from the above discussion,
since the penetration of the envelope absorbs energy in order to lift the
dredged-up material into the envelope (see also Paczy\'nski 1977).
Dredge-up also leads to lower surface radii.

\subsection{Influence of convection prescription}
\label{Sect:convection prescription}

While the dredge-up characteristics are found to be rather insensitive to the
extra-mixing parameters, a dependency of those characteristics on the convection
prescription {\it in} the envelope should be discussed.
Higher values of $\alpha_{mlt}$ are known to favor deeper convective envelopes
(Wood 1981). The envelope would thus reach the H-depleted layers sooner and lead,
for an otherwise similar dredge-up rate, to more C-rich material being
dredged-up.
Moreover, the dredge-up rate itself should also be affected by the convection prescription.
The analysis presented in Sect.~\ref{Sect:dredge-up process} stresses on the role
of the thermal response of the envelope in the dredge-up process. The thermal
time-scale of the envelope depends
on its thermal structure, which is imposed by the surface
conditions {\it and} the $\alpha_{mlt}$ parameter. We thus expect the dredge-up
rate to be also dependent on $\alpha_{mlt}$. A test calculation
on the 19th afterpulse of the \mass{3} computed with $\alpha_{mlt}=2.2$
leads to $r_{dup}=\mmass{6.9\times 10^{-5}}$~/yr with $M_c$=\mass{0.65}.
This has to be compared with a predicted
dredge-up rate of \mass{5.7\times 10^{-5}}~/yr at that core mass in models with
$\alpha_{mlt}=1.5$ (from Eq.~\ref{Eq:r_dup}). Models with $\alpha_{mlt}=2.2$ thus
predict higher dredge-up rates by about 16\% than those in models with
$\alpha_{mlt}=1.5$.

\subsection{Comparison with other published works}
\label{Sect:comparison}

The calculations presented in Sects.~\ref{Sect:no overshooting} and
\ref{Sect:overshooting} lead to a simple conclusion about AGB models using the
Schwarzschild criterion: models computed without any extra-mixing do not lead to
dredge-up while those using an extra-mixing procedure lead to efficient dredge-up.

How does this compare to model calculations published in the literature? The
answer to that question requires the knowledge of the precise numerical
procedures used by each AGB modeler in their code. Such information is often not
available in the literature. For this reason -- and because it is anyway outside
the scope of this paper to do so -- it is impossible to compare all published
models. Some of them are however discussed below.

  Many AGB calculations of the last five years do not report the
occurrence of the 3DUP (Vassiliadis \& Wood 1993, Wagenhuber \& Weiss 1994,
Bl\"ocker 1995, Forestini \& Charbonnel 1997).
This is consistent with their working hypothesis of
the Schwarzschild criterion without extra-mixing.

  Several codes use an extra-mixing procedure based on a purely numerical
technique. Boothroyd and Sackmann (1988a), for example, mix the radiative layer
adjacent to the convective zone {\it before} checking for its stability. If the
$\nabla_{rad}/\nabla_{ad}$ ratio of that (formally stable) layer is found to be
greater than one after mixing, then they declare it convective. As this mixing
procedure is performed once per model calculation, it allows a numerical
propagation of the convective envelope during the 3DUP. This might be the reason
for their success in obtaining low-mass carbon stars (Boothroyd \& Sackmann
1988b). A similar technique was already used by Paczy\'nski (1977). This last
author obtained dredge-up in \mass{8} AGB models with such a numerical
`overshoot' procedure, while no dredge-up resulted in models not including that
overshoot. The use of extra-mixing procedures based on a purely numerical
technique suffers from at least two shortcomings. First, they assume
instantaneous mixing of the layer(s) added to the envelope. Besides being
unphysical, it may lead to convergence difficulties. Second, they would probably
depend on the numerical accuracies of the models such as the time-step and
mesh re-zoning. As a result, such models may or may not lead to efficient
dredge-up.

  Herwig et al. (1997) use an overshooting algorithm which assumes an
exponentially decreasing bubble velocity field in the extra-mixing region. Their
algorithm is very similar to mine, and enables a direct comparison of their
predictions with mine. These are reported in Fig.~\ref{Fig:lambda}{\sl
(b)} ($+$ signs).
Interestingly, they find higher dredge-up efficiencies than I do at a given
stellar luminosity. We note that they use $\alpha_{mlt}=1.7$ in their
calculations. This could explain part of the differences. We further note that their
core masses are lower by about \mass{0.01} than mine at a given pulse number,
which could imply a dependence of the dredge-up efficiencies on the initial
conditions at the onset of the AGB.

  Finally, the recent work of Straniero et al. (1997) should be mentioned. These
authors obtain dredge-up in models of 1.5 and \mass{3} stars without using
any extra-mixing procedure in their calculation. This seems in contradiction with
my conclusions. Private discussions with some of the authors in
Straniero et al. (1997) seemed to confirm that they did not use {\it any}
extra-mixing in their model calculations. They explain their results by
the higher temporal and spatial resolution of their models (see their paper for
more details). This conclusion, too,
is not supported by my calculations reported in Sect.~\ref{Sect:no
overshooting}. While I do not understand their results,
I cannot push the analysis much further.
In any case,
the dredge-up efficiencies found in their models are lower than those predicted from
my calculations including extra-mixing.
Their predictions are reported in Fig.~\ref{Fig:lambda}{\sl (b)} ($\times$ signs).

\section{Third dredge-up laws}
\label{Sect:dredge-up laws}

\subsection{The dredge-up process}
\label{Sect:dredge-up process}

The important conclusion that the dredge-up rate is insensitive to
the extra-mixing parameters can be understood from the following considerations.
The AGB star is basically formed by an
$e^-$-degenerate core of 0.10-0.15~R$_\odot$ surrounded by an extended H-rich
envelope. As dredge-up proceeds, the outermost
H-depleted layers of the core become part of the envelope and expand
[which translates into a negative gravothermal energy production
in those layers, see Fig.~\ref{Fig:gradov}{\sl (c)}]. This results in an increase of
the potential energy of those layers engulfed in the convective envelope,
which must be supplied by the luminosity
provided by the inner layers. {\it The maximum dredge-up rate is then determined
by the thermal relaxation time-scale of those layers expanding into the envelope
during the dredge-up.}

Let us estimate that dredge-up rate from simplified arguments. The potential
energy $E_{pot}$ of a mass $\Delta m$ located at the H-He discontinuity is given by
\begin{equation}
\label{Eq:potential energy}
  E_{pot}=-G\;\frac{\Delta m\; M_c}{R_c},
\end{equation}
where $M_c$ is the mass of the H-depleted core and $R_c$ its radius
(i.e. $R_c=R_{env,b}$ during dredge-up). The
energy necessary to lift the dredged-up material into the envelope is
provided by the luminosity at the core edge, $L_c$.
The thermal time-scale $t_{dup}$ for the dredge-up is then
\begin{eqnarray}
  t_{dup} & = & \frac{E_{pot}}{L_c} \nonumber \\
          & = & 3.12\times 10^7\;\frac{M_c}{R_c\; L_c} \;\Delta m.
\label{Eq:t_dup theoretical}
\end{eqnarray}
All quantities in this section are expressed in solar units
and years. The dredge-up rate can then be estimated by
\begin{eqnarray}
  r_{dup} & = & \frac{\Delta m}{t_{dup}} \nonumber \\
          & = & 3.2\times 10^{-8}\;\frac{R_c\; L_c}{M_c}.
\label{Eq:r_dup theoretical}
\end{eqnarray}
Typical values for the \mass{3} star are $M_c\simeq \mmass{0.66}$
[Fig.~\ref{Fig:pulse19}{\sl (a)}], $R_c\simeq \mrsun{0.12}$
[Fig.~\ref{Fig:gradov}{\sl (d)}]
and $L_c\simeq \mlsun{10^{4}}$ [Fig.~\ref{Fig:epsgrav}{\sl (b)}].
Equation \ref{Eq:r_dup theoretical} then gives $r_{dup} \simeq
\mmass{5.8\times 10^{-5}}$/yr.
This estimate of the dredge-up rate, while giving only an order of magnitude,
convincingly supports the rate of \mass{4.2\times 10^{-5}}/yr obtained in the
full evolutionary models (Sect.~\ref{Sect:overshooting}). It is also consistent with the results
obtained by Iben (1976), who estimated the dredge-up rate to be
about \mass{10^{-5}}/yr in a \mass{7}
star with $M_c=0.95-\mmass{0.96}$, and with those of Paczy\'nski (1977), who found
\mass{5.2\times 10^{-5}}/yr in a \mass{8} star with $M_c=\mmass{0.8}$.

The question of why the dredge-up rate is independent of the extra-mixing
parameters can further be understood by comparing the time-scale of the
thermal readjustment of the envelope with the typical time-scale of convective bubbles to cross the H-He
transition zone. The velocity of the convective bubbles as they approach the H-He
discontinuity is of the order of \rsun{10}/yr, which translates in terms
of mass to about \mass{10^{-1}}/yr (from Fig.~\ref{Fig:gradov}{\sl(d)}). This
rate is much higher than the dredge-up rate of \mass{\sim 10^{-5}}/yr established
above. The deepening of the envelope into the H-depleted layers is thus primarily
fixed by the thermal relaxation time-scale of the envelope, rather than by the
speed of the convective bubbles penetrating into the H-depleted layers.

\subsection{Dredge-up characteristics}
\label{Sect:dredge-up characteristics}

Equation~\ref{Eq:r_dup theoretical} reveals a formal dependence of the dredge-up
rate on $R_c$, $M_c$ and $L_c$.

Those three stellar parameters are not independent of each other. Refsdal \&
Weigert (1970) and Kippenhahn (1981), for example, show from homology considerations
that most red giant star properties
are functions of the H-depleted core mass and radius. As the burning shell
advances, the structure of the intershell layers evolves, to first approximation,
like homologous transformations. Let us consider two times $t$ and $t'$
characterized by core masses, radii and luminosities of $M_c$, $R_c$ and $L$,
respectively, at time $t$, and $M'_c$, $R'_c$ and $L'$, respectively, at time
$t'$. Then the homology transformations applied to a \mass{\sim 3} AGB star lead
to (Kippenhahn 1981, Herwig et al. 1998)
\begin{equation}
\label{Eq:McRcL}
  \frac{L'}{L} \propto \left(\frac{M'_c}{M_c}\right)^2\;\left(\frac{R'_c}{R_c}\right)^{-1}.
\end{equation}
The dredge-up rate $r'_{dup}$ at time $t'$ can then easily be evaluated from that at time $t$,
$r_{dup}$. Equations~\ref{Eq:r_dup theoretical} and ~\ref{Eq:McRcL} lead
to
\begin{equation}
\label{Eq:r_dup relation}
  \frac{r'_{dup}}{r_{dup}} \propto \frac{M'_c}{M_c}.
\end{equation}

The homology transformations thus suggest a {\it linear relation between
the dredge-up rate and the core mass}.

   In standard AGB calculations (i.e. without dredge-up), $R_c$
is a linear function of $M_c$ (Eq.~\ref{Eq:McenvRb linear} for our
\mass{3} star). Equation~\ref{Eq:McRcL} then recovers
the classical linear $M_c$-$L$ relation. In that case, $r_{dup}$
is a linear function of {\it either } $M_c$, $R_c$ or $L$.
The dredge-up predictions presented in
Sect.~\ref{Sect:overshooting} essentially obey these rules
(Table~\ref{Tab:dredge-ups}), since the dredge-up calculations
have been performed on {\it selected} afterpulses of the standard
\mass{3} models and do not include the feedback of the dredge-ups on the AGB
evolution.

  When the feedback of dredge-up on the structural evolution is taken into
account, the core radius is no more a linear relation of the core mass,
and the linear $M_c$-$L$ does not hold any more (Herwig et al. 1998). It must be
replaced by a $M_c$-$R_c$-$L$ relation as suggested by Eq.~\ref{Eq:McRcL}.
Equation~\ref{Eq:r_dup relation}, however, reveals that the dredge-up
rate keeps a linear dependency on $M_c$ even in models experiencing dredge-up.

  The implications of the linear $M_c$-$r_{dup}$ relation on the core mass
evolution are analyzed in Sect.~\ref{Sect:WD mass}.

\section{Dredge-up and the formation of white dwarfs}
\label{Sect:AGB evolution}

  The dredge-up laws established in Sects.~\ref{Sect:overshooting}
and \ref{Sect:dredge-up laws} allow an approximate study of the structural
evolution of AGB stars without actually performing full AGB model
calculations. This is the so called `synthetic' AGB calculation technique.

  Such a study is outside the scope of the present paper. However, a
preliminary analysis of the synthetic evolution of a \mass{3} star with dredge-up is
presented in Sects.~\ref{Sect:structural evolution} and \ref{Sect:C star}
using simplified assumptions.
But before doing that, the implications of the linear dependence of the dredge-up rate
on the core mass is first discussed in Sect.~\ref{Sect:WD mass}.

\subsection{Core mass evolution and the formation of white dwarfs}
\label{Sect:WD mass}

The analysis presented in Sect.~\ref{Sect:dredge-up laws} supports a linear
relation between $r_{dup}$ and $M_c$. I assume that such a linear
relation remains valid between $\lambda$ and $M_c$ even
when the feedback from the dredge-ups on the structural AGB evolution is taken into account.

Such a linear relation between $\lambda$ and $M_c$ has an important
consequence on the core mass reached at the end of the AGB evolution.
Equation~\ref{Eq:lambda} predicts that $\lambda$ reaches unity at $M_c^*=\mmass{0.71}$.
It is easy to see that
the feedback between $\lambda$ and $M_c$ leads to the evolution of $M_c$ towards
an asymptotic value
$M_c\!=\!M_c^*$. Indeed, if $M_c\!<\!M_c^*$, then $\lambda\!<\!1$ and the core mass
increases from one pulse to the next. If $M_c\!>\!M_c^*$, then $\lambda\!>\!1$ and the core
mass {\it decreases} from one pulse to the next. At $M_c\!=\!M_c^*$,
the amount of material dredged-up from the core during the 3DUP
equals that added to the core by H-burning during the interpulse phase, and
$M_c$ remains constant from one pulse to the next.
We thus expect $\lambda$ to level off at unity, and $M_c$ to reach $M_c^*$.
This puts an upper limit to the mass of white dwarf remnants (which is
\mass{0.71} for our \mass{3} star), and helps to
constrain the predicted initial-final mass relation for white dwarfs.

\subsection{Structural evolution}
\label{Sect:structural evolution}

\begin{figure}
  \resizebox{\hsize}{!}{\includegraphics{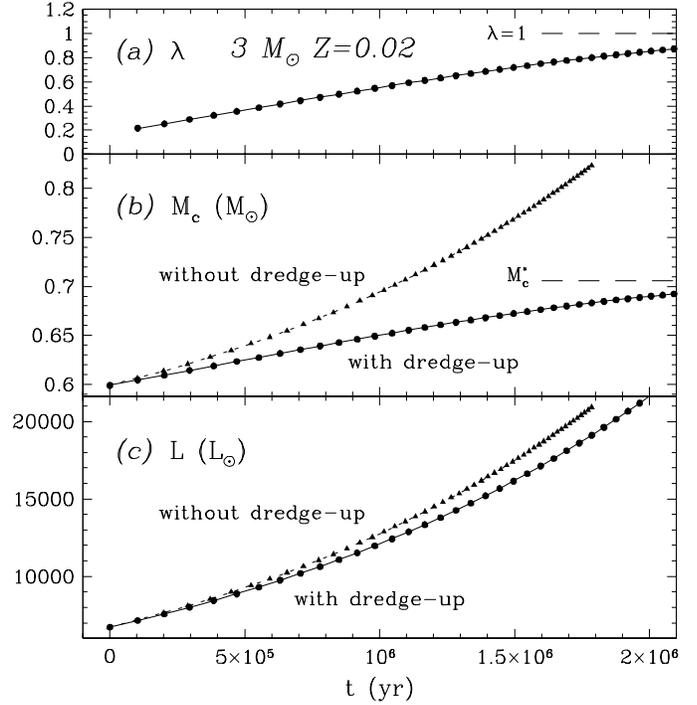}}
  \caption{Predictions from synthetic calculations for a \mass{3}
           AGB star with (filled circles) and without (filled triangles)
           dredge-up.
           {\sl (a)} Dredge-up efficiency, {\sl (b)} core mass,
           and {\sl (c)} surface luminosity as a function of time.
           The first displayed pulse (at time $t=0$) corresponds to the
           10th pulse of the standard \mass{3} models.
           The horizontal dashed line in {\sl (b)} shows
           the asymptotic core mass predicted for models experiencing
           dredge-up.
          }
  \label{Fig:synth}
\end{figure}

 According to Eq.~\ref{Eq:r_dup}, dredge-up begins to operate in our \mass{3}
star around the 10th pulse, at \mass{M_c=0.599}. This is where we begin our synthetic
evolution. We assume that the star has reached its asymptotic regime,
which is a good enough assumption (see Fig.~\ref{Fig:relations}) for our
purposes.

 The evolution of $M_c$ is governed, on the one hand, by its increase
$\Delta M_{c,nuc}$ due to hydrogen burning during the interpulse.
For a solar metallicity star, $\Delta M_{c,nuc}$ is given by
(all quantities in this section are expressed in solar units and in years)
\begin{eqnarray}
  \Delta M_{c,nuc} & =      & f_L \times 1.37\times 10^{-11}\; L_H\; \times \Delta t_{interp} \nonumber \\
\label{Eq:Delta M_nuc}
                   & \simeq & f_L \times 1.37\times 10^{-11}\; L\; \times \Delta t_{interp},
\end{eqnarray}
where the luminosity $L_H$ of the H-burning shell is approximated by the surface
luminosity (at the time of the maximum extension of the next pulse). The factor
$f_L$ is introduced in order to account for the fact that the mean surface luminosity
during the interpulse is actually lower than that at the the next pulse. From the
standard model calculations, we find $f_L=0.72$.
 After each pulse, on the other hand, $M_c$ decreases as a result of the
3DUP. The amplitude of this decrease is given by
$M_{dup}=\lambda\times\Delta M_{c,nuc}$,
where $\lambda$ is estimated from Eq.~\ref{Eq:lambda}.
The resulting net increase in $M_c$ from one pulse to the next is then
\begin{equation}
\label{Eq:Delta Mc}
  \Delta M_c = (1-\lambda)\times \Delta M_{c,nuc}.
\end{equation}

In model calculations without dredge-up, the change in luminosity $\Delta L$
from one pulse to the next in the asymptotic regime is given by
Eq.~\ref{Eq:McL linear},
while the change in the core radius $\Delta R_c$ is given by
Eq.~\ref{Eq:McenvRb linear}.
In the presence of dredge-up, those relations must be modified. For $\Delta L$,
a dependence on $(\Delta M_c)^2/\Delta R_c$ is suggested from Eq.~\ref{Eq:McRcL},
and we adopt
\begin{equation}
\label{Eq:Delta L}
  \Delta L = - 80000\; \frac{(\Delta M_c)^2}{\Delta R_c}
\end{equation}
For $\Delta R_c$, we make the assumption that the {\it time} evolution of $R_c$ remains
unaffected by the dredge-ups. Equations~\ref{Eq:Mcdt linear} and \ref{Eq:McenvRb linear}
then determine $R_c$ as a function of time $t$.

\begin{figure}
  \resizebox{\hsize}{!}{\includegraphics{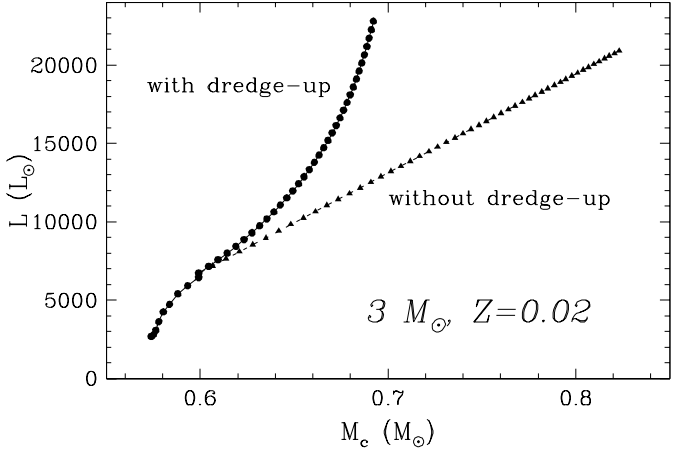}}
  \caption{Luminosity as a function of core mass predicted by the synthetic
           calculations for a \mass{3} star when dredge-up is (filled circles) or is not
           (filled triangles) taken into account.
          }
  \label{Fig:synthMc}
\end{figure}

Finally, the interpulse duration is assumed not to be altered by the dredge-up
episodes, and is given by Eq.~\ref{Eq:Mcdt}.

The predictions of our synthetic calculations are displayed in Figs.~\ref{Fig:synth}
(time evolution) and \ref{Fig:synthMc} ($M_c$-$L$ evolution). As expected from
the discussion in Sect.~\ref{Sect:WD mass}, $\lambda$ increases towards unity and
the core mass towards the asymptotic value of $M_c^*=\mmass{0.71}$. The
luminosity, however, increases continuously. This results from the fact that $L$
depends not only on $M_c$ but also on $R_c$ according to Eq.~\ref{Eq:McRcL}. Had
we used a linear $M_c-L$ relation, then $L$ would have evolved towards an
asymptotic value (of about \lsun{14000} for our \mass{3} star) in a similar way
as does $\lambda$. The non-linearity of $L$ with $M_c$ is made clear in
Fig.~\ref{Fig:synthMc}.

  The assumptions underlying those synthetic calculations certainly are too
simplistic. For example, the feedbacks of dredge-up on $\lambda$, $\Delta
t_{interp}$ or $\Delta R_c$ have been neglected, and relations~\ref{Eq:McRcL} and
\ref{Eq:r_dup relation}
need to be confirmed by model calculations following the dredge-ups all
along the AGB evolution. Yet, the main conclusions, such as
the existence of a limiting $M_c^*$ towards which the core mass evolves
asymptotically, should qualitatively be correct.

\paragraph{Effects of mass loss}  A last word about the effects of mass-loss.
The amplitude of mass loss depends
mainly on the surface radius $R$ and effective temperature $T_{eff}$ (and thus on
$L$). Those are
expected to be modified by the dredge-ups (Fig.~\ref{Fig:pulse19}). However, the time evolution
of $L$ [Fig.~\ref{Fig:synth}{\sl (c)}] reveals that they are not as much affected as
is $M_c$. As a result, mass loss is expected to be higher
in models with dredge-up than in those without, at a given $M_c$.
In other words, mass loss will
eject the AGB's envelope at a much lower value of $M_c$ in the presence of
dredge-up.
For example, let us suppose that a superwind suddenly ejects all the
envelope of the \mass{3} star at \lsun{L=15000}. Figure~\ref{Fig:CO} then
predicts that the mass of the white dwarf's
remnant would be \mass{\sim 0.66} in the presence of dredge-ups. This is much
lower than the predicted white dwarf's mass of \mass{0.73} predicted in the absence of dredge-up.
Mass loss combined with dredge-ups thus further helps to constrain the mass of white dwarf remnants.

\subsection{Carbon star formation}
\label{Sect:C star}

\begin{figure}
  \resizebox{\hsize}{!}{\includegraphics{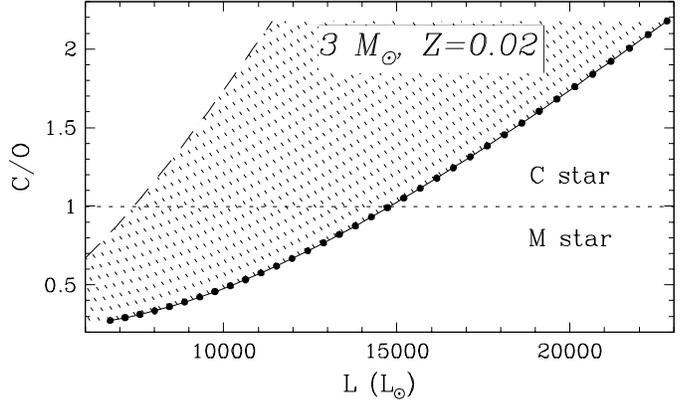}}
  \caption{Same as Fig.~\ref{Fig:synthMc}, but for the surface C/O ratio
           as a function of surface luminosity at maximum pulse extension
           (solid line) or at the luminosity dip during the interpulse phase
           (taken here to be half of the luminosity at maximum pulse extension,
           dashed line).
           The dotted line divides the regions where C stars (C/O$>$1)
           and M stars (C/O$<$1) are.
           The hatched area indicates the range of luminosity predicted for
           given C/O ratios.
          }
  \label{Fig:CO}
\end{figure}

Let us now follow through our synthetic calculations the carbon abundance predicted at
the surface of a \mass{3} star.
The amount of carbon dredged-up to the surface, $M^C_{dup}$, is
given by
\begin{equation}
\label{Eq:delta M(C)}
  M^C_{dup}= X_p(\mchem{C}{12})\;\times\; M_{dup},
\end{equation}
where $X_p(\mchem{C}{12})$ is the \chem{C}{12} mass fraction
left over by the pulse in the intershell layers. From the values reported
in Table~\ref{Tab:dredge-ups}, a mean value of
$X_p(\mchem{C}{12})=0.21$ can be assumed. The initial
\chem{C}{12} and \chem{O}{16} mass fractions in the envelope are taken from our
standard \mass{3} star at the beginning of the TP-AGB phase, and are equal to
$2\times 10^{-3}$ and $9.7\times 10^{-3}$, respectively.
The resulting evolution of the surface C/O number ratio is shown in
Fig.~\ref{Fig:CO}. It is seen that a \mass{3} star satisfying Eq.~\ref{Eq:lambda}
becomes a C star after about twenty dredge-up episodes,
at $L\simeq \mlsun{15000}$. Furthermore, it is
known that the luminosity decreases by a factor of almost two during about
20\% of the interpulse phase.
This means that the star displayed
in Fig.~\ref{Fig:CO} could be observed as a C-star at luminosities as low as
\lsun{7500}.

I should stress that the effects of mass loss have been neglected in those
synthetic calculations, being outside the scope of this article. Mass loss would
decrease the dilution factor of \chem{C}{12} into the envelope, but would probably also
decrease the dredge-up efficiency. Those effects should be considered in a
future, more detailed, study of C star formation.

\section{Conclusions}
\label{Sect:conclusions}

The analysis presented in this paper addresses several questions
regarding the occurrence of 3DUP in AGB star models, the laws governing that
phenomenon, and some of its implications on the structural and chemical evolution
of those stars. The analysis is based on model calculations of a \mass{3} star
using the Schwarzschild criterion and performed either with or without extra-mixing.

\paragraph{Modeling dredge-up in AGB stars}

 The use of a local criterion to
delimit convection borders, such as that of Schwarzschild, and without any
extra-mixing is shown to lead to an unphysical situation which prevents the
deepening of the convective envelope into the H-depleted regions, and which thus
also prevents the
occurrence of the 3DUP (Sect.~\ref{Sect:boundaries}). The Schwarzschild layer
is undefined in those models due to the development of a discontinuity
in the hydrogen abundance profile at the bottom of the convective envelope. Model
calculations using no extra-mixing confirm the failure to obtain dredge-up in
those conditions (Sect.~\ref{Sect:no overshooting}). That conclusion is found to
be independent of the mixing length parameter, stellar mass, or numerical space
and time resolution of the models. Models using the Schwarzschild criterion with
no extra-mixing are thus inadequate to describe the 3DUP phenomenon.

Models of the same \mass{3} star but using extra-mixing, on the other hand, lead
to efficient dredge-ups from the 11th pulse on (Sect.~\ref{Sect:overshooting}).
This results directly from the
unstable character of the lower boundary of the envelope against extra-mixing
(Sect.~\ref{Sect:unstable boundary}). The calculations further reveal that the
dredge-up characteristics are insensitive to the extra-mixing
parameters (such as the extra-mixing extent and efficiency). This important
conclusion is most welcomed since nothing is known yet about the characteristics
of the extra-mixing expected to occur in AGB stars. It enables, in particular, to
study the dredge-up characteristics with some confidence and without worrying
about the extra-mixing parameters.

Although the dredge-up predictions are rather insensitive to the extra-mixing
parameters, a proper handle of the extra-mixing procedure is essential in order to obtain
reproducible predictions. In particular, a purely numerical extra-mixing can
lead to model-dependent predictions (Sect.~\ref{Sect:comparison}; see also
Mowlavi 1999a). The use of instantaneous mixing in the extra-mixing region can
also lead to convergence difficulties. The calculations presented in this paper
use a diffusive overshooting with a decreasing bubble velocity field in the
extra-mixing region. It
leads to smooth chemical abundance profiles and to a proper location of the
Schwarzschild layer.

\paragraph{The dredge-up process}

The insensitivity of the dredge-up predictions on the extra-mixing parameters
results from the fact that the dredge-up rate is limited by the
time scale of the thermal readjustment of the envelope (Sect.~\ref{Sect:dredge-up process}). As
dredge-up proceeds, H-depleted matter is lifted from the core into the envelope
and its thermal state must relax to that of the envelope (see also Mowlavi 1999b).
The dredge-up rate is estimated to \mass{10^{-5}-10^{-4}}/yr from
model calculations (Sect.~\ref{Sect:overshooting}). This is also what is expected
from an analytical estimation
using simplified arguments of the physics involved in the dredge-up process
(Sect.~\ref{Sect:dredge-up process}).

\paragraph{Dredge-up laws}

The model calculations performed with extra-mixing on selected afterpulses
of the standard \mass{3} star predict linear relations both between the dredge-up
rate and core mass (Eq.~\ref{Eq:r_dup}) and between
dredge-up efficiency and core mass (Eq.~\ref{Eq:lambda}). In those calculations
where the feedback of dredge-up on the AGB evolution is not taken into account,
the dredge-up rate and efficiency further depend linearly on the stellar luminosity.

When the feedback of dredge-up on the AGB evolution is taken into account, the
analysis presented in Sect.~\ref{Sect:dredge-up laws} suggests that the dredge-up
rate keeps its linear dependence on the core mass. This conclusion is valid if
the $M_c-R_c-L$ relation is of the form of Eq.~\ref{Eq:McRcL}, $R_c$ being the radius
of the H-depleted core. In that case, we expect the dredge-up efficiency to level
off at $\lambda=1$, at which point the core mass does not increase anymore from
one pulse to the next. The asymptotic core mass of the \mass{3} star model
presented in this paper is \mass{M_c^*=0.71} (Sect.~\ref{Sect:WD mass}).

\paragraph{Consequences on the AGB evolution}

The synthetic calculations performed in Sect.~\ref{Sect:AGB evolution} confirm the
asymptotic evolution of the dredge-up efficiency towards unity and of the core mass
towards $M_c^*$. This helps to constraint the initial-final mass relation of
white dwarfs, $M_c^*$ giving an upper limit for the mass f the white dwarf
remnant.

The surface luminosity is not as much affected by the dredge-ups as is $M_c$.
This results from the adopted $M_c-R_c-L$ relation combined
with the continuous decrease of $R_c$.
The non-validity of the $M_c-L$ linear relation, assumed in Eq.~\ref{Eq:Delta L},
is made evident in Fig.~\ref{Fig:synthMc}.
The increase in $L$ with time predicted by the synthetic calculations is only
slightly less steep than the $L(t)$ increase predicted in the standard
calculations without dredge-up (Fig.~\ref{Fig:synth}).

Mass loss can further constrain the initial-final mass relation of white dwarfs,
because it depends on the luminosity through the stellar radius and effective temperature
(Bowen \& Willson 1991), and because the luminosity increases with time despite the
leveling of $M_c$.
We thus expect mass loss
[or superwind(s)] to peal off the AGB's envelope at a lower core mass than in
models not experiencing dredge-up. For example, if a superwind is assumed to
eject all the envelope when the star reaches \lsun{L=15000}, the \mass{3} star models
with dredge-up predict the formation of a white dwarf remnant of \mass{0.66}.
This has to be compared with a predicted remnant of \mass{0.73} in the absence of
dredge-up. The synthetic calculations presented in Sect.~\ref{Sect:AGB evolution} are,
however, preliminary and based on simplified assumptions. Future investigations
along these lines should refine the quantitative predictions given in this paper.

Finally, a \mass{3} carbon star is predicted to form after about twenty pulses
experiencing dredge-up (i.e. at about the 25th-30th pulse on the AGB), at $L\simeq
\mlsun{15000}$. Taking into account the luminosity dip experienced during about
20\% of the following interpulse time, such a C star could be observed at
luminosities as low as \lsun{7500}.

\noindent {\small{\it Acknowledgments.}  
I thank Drs. A. Jorissen and G. Meynet for their careful reading of the first
version of this article, and Dr. G. Meynet for many useful discussions.

\end{document}